\documentclass[12pt]{article}
\usepackage{enumerate,amssymb}
\renewcommand{\a}{\alpha}
\renewcommand{\b}{\beta}
\newcommand{\g}{\gamma}           
\renewcommand{\d}{\delta}

\newcommand{\la}{\lambda}

\newcommand{\s}{\sigma}           
         \newcommand{\T}{\Theta}
\newcommand{\f}{{\phi}}

\newcommand{\be}{\begin{equation}}
\newcommand{\ee}{\end{equation}}
\newcommand{\eqn}[1]{\label{#1}\end{equation}}

\newcommand{\bea}{\begin{eqnarray}}
\newcommand{\eea}{\end{eqnarray}}
\newcommand{\eqan}[1]{\label{#1}\end{eqnarray}}

\newcommand{\ba}{\begin{array}}
\newcommand{\ea}{\end{array}}

\newcommand{\nn}{\nonumber}

\begin{document}

\begin{center}
{\bf   Inclusion of Yang-Mills Fields in String Corrected Supergravity}\\[14mm]

S. Bellucci\\

{\it INFN-Laboratori Nazionali di Frascati\\
Via E. Fermi 40, 00044 Frascati, Italy\\ mailto: bellucci@lnf.infn.it}\\[6mm]

D. O'Reilly\\

{\it Department of Mathematics, The Graduate School and University
Center\\365 Fifth Avenue
New York, NY 10016-4309\\ mailto: doreilly@gc.cuny.edu and dunboyne@vzavenue.net}\\[6mm]

\end{center}
\vbox{\vspace{3mm}}

\begin{abstract}

We consistently incorporate Yang Mills matter fields into string
corrected (deformed),
 D=10, N=1 Supergravity.
 We solve the Bianchi identities within the framework of the modified beta function favored constraints to second order in the string slope parameter $\g$ also including the Yang Mills fields.
  In the torsion, curvature and H sectors we find that a consistent solution is readily obtained with a Yang Mills modified supercurrent  $A_{abc}$.
  We find a solution in the F sector following our previously developed method.
 \vbox{\vspace{1mm}}

\end{abstract}

\vbox{\vspace{3cm}}

PACS number: 04.65.+e

\newpage

%%%%%%%%%%%%%%%%%%%%%%%%%%%%%%%%%%%%%%%%%%%%%%%%%%%%%%%%%%%

\section{Introduction}

String corrections to quantum field theories are believed to
contribute Gauss Bonnet terms to the action. This invariant has
been studied in many different scenarios, for example Gauss-Bonnet
modified cosmology \cite{10}. These terms are introduced by hand
into the quantum gravity models. However it is also known that
Gauss-Bonnet terms occur naturally in the context of string
corrected gravity \cite{5}. The leading order corrections in terms
of the string tension parameter, $\g$ contain these invariants.
Hence we are motivated by this and by other reasons to study
string deformed supergravity, with D=10, N=1 as the low energy
limit of string theory. Here we wish to include Yang Mills fields
in the non minimal theory. As several of the terms are extremely
lengthy we avoid writing them explicitly. In a future work we will
explore the simplified bosonic case.

Some years ago a scenario was developed to construct a manifestly
supersymmetric theory of string corrected supergravity
\cite{1,2,3,4,5}. This involved incorporating the Lorentz
Chern-Simons super form, $X^{LL}$ into the geometry of D=10 N=1
supergravity. It was initially successful at first order in $\g$,
\cite{5}. It ran into difficulties and controversy at second
order, \cite{6}. It was suggested that at second order the torsion
$T_{\a\b}{}^{g}$ should be modified to include the so called X
tensor, \cite{1}. A search for the X tensor ensued and a candidate
was proposed in \cite{12} which was shown to allow for the
solution of the Bianchi identities in the H sector and also the
torsion and curvature sectors.

In this paper we show that a simple modification of the X tensor
Ansatz allows also for the inclusion of matter fields in the H,
torsion and curvature sectors. We show that the assumption
$F^{(2)}{}_{\a\b}=0$ does not allow for a solution. We propose a
candidate that does allow for a solution. Prior to finding the $X$
tensor also by way of an Ansatz, in \cite{12}, the search for a
second order solution consisted of systematically studying the
table of irreducible representations, \cite{13}. However the task
proved formidable because of the number of unknown quantities. It
was eventually shown that the form given in equation (33) of the
first reference \cite{12} for $T^{(2)}{}_{\a\b}{}^{g}$, satisfying
the torsion identity at dimension one half, thereafter allowed
ultimately for a consistent solution in the torsion and curvature
sectors.

In reference \cite{12}, we found a solution to D=10, N=1
supergravity to second order in the string slope parameter with
the modified tensor \bea  G_{ADG}~= H_{ADG}~+~\g Q_{ADG}\eea Here
we extend this as follows, \bea G_{ADG}~= H_{ADG}~+~\g Q_{ADG}+ \b
Y_{ADG} \eea Here $\b$ is the Yang-Mills coupling constant. Matter
fields will not have consequences for the appearance of the
Gauss-Bonnet term however, but we wish to construct a complete
model.

\section{The Solution}

 The Bianchi identities in Superspace are as follows:
\bea[[\nabla_{[A},\nabla_{B}\},\nabla_{C)}\}=0   \eea Here we have
extended the commutator in \cite{12} to include the Yang-Mills
field strengths, $F_{AB}^{I}$ \bea
[\nabla_{A},\nabla_{B}\}=T_{AB}{}^{C}\nabla_{C}+\frac{1}{2}R_{AB}{}^{de}M_{ed}+iF_{AB}^{I}t_{I}
\eea
 The $t_{I}$ are the generators of the Yang-Mills gauge group. For notation convenience we drop
 the group index $I$. For notational brevity we also write
\bea
R_{ABde}=R^{(0)}{}_{ABde}+R^{(1)}{}_{ABde}+R^{(2)}{}_{ABde}+....\eea
and \bea
T_{AD}{}^{G}=T^{(0)}{}_{AD}{}^{G}+T^{(1)}{}_{AD}{}^{G}+T^{(2)}{}_{AD}{}^{G}+...\eea
Where the numerical superscript refers to the order of the
quantity. A quantity which re-occurs is the following \bea
\Omega^{(1)}{}_{gef}=L^{(1)}{}_{gef}-\frac{1}{4}A^{(1)}{}_{gef}\eea
and its spinor derivative which we denote simply as \bea
\Omega^{(1)}{}_{\g
gef}=\nabla_{\g}[L^{(1)}{}_{gef}-\frac{1}{4}A^{(1)}{}_{gef}] \eea
We evaluate this in appendix. Reference \cite{5} began with the
conventional constraints as follows \bea
i\s_{b}{}^{\a\b}T_{\a\b}{}^{a}=16\d^{a}{}_{b}\nn\\
\s_{abcde}{}^{\a\b}T_{\a\b}{}^{a}=0\nn\\ \s_{a}{}^{\a\b}T_{\a
\b}{}^{a}=0\nn\\ T_{\a [bc]}=0 \nn\\T_{abc}=\frac{1}{6}T_{[abc]}\nn\\
T_{a \a}{}^{\b}=
\frac{1}{48}\s_{a}{}_{\a\la}\s^{pqr}{}^{\b\la}A_{pqr}\nn\\~~ \eea

 Using these constraints then led to the first order solution. That is $G^{(1)}{}_{ABC}$, $T^{(1)}{}_{AB}{}^{C}$,
 and $R^{(1)}{}_{ABde}$ were found, along with additional modified constraints. In \cite{12} we found that no modification
 to the condition $H_{\a\b\g}=0$ was necessary. We continue with this assumption.

We have the H sector Bianchi identities with Yang-Mills fields as
follows \cite{5}:

\bea
\frac{1}{6}\nabla_{(\a|}H_{|\b\g\d)}~-~\frac{1}{4}T_{(\a\b|}{}^{M}
H_{M|\g \d)}~=~-\frac{\g}{4}R_{(\a\b|ef}R_{|\g \d)}{}^{ef}
-\frac{\b}{4}F_{(\a\b|}{}^{I}F_{|\g \d)}{}^{I}\eea

\bea
\frac{1}{2}\nabla_{(\a|}H_{|\b\g)d}~-~\nabla_{d}H_{\a\b\g}~-~\frac{1}{2}T_{(\a\b|}{}^{M}H_{M|\g)d}~
+~\frac{1}{2}T_{d(\a|}{}^{M}H_{M|\b\g)}~\nn\\=~-\g
R_{(\a\b|ef}R_{|\g)d}{}^{ef}-\b F_{(\a\b|}{}^{I}F_{|\g)d}{}^{I} \eea

 \bea~\nabla_{(\a|}H_{\b)
cd}~+~\nabla_{[c|}H_{|d]\a\b}~-~T_{\a\b}{}^{M}H_{Mcd}~
-~T_{cd}{}^{M}H_{M\a\b}~-~T_{(\a|[c|}{}^{M}H_{M|d]|\b)}~=~\nn \\
~-\g[2R_{\a\b ef}R_{cd}{}^{ef}~
+~R_{(\a|[c|}{}_{ef}R_{|d]\b)}{}^{ef}]-\b[2F_{\a\b}{}^{I}F_{cd}{}^{I}~
+~F_{(\a|[c|}{}^{I}F_{|d]\b)}{}^{I}]\eea We must also solve the
torsions and curvatures \bea
T_{(\a\b|}{}^{\la}T_{|\gamma)\la}{}^{d}~-~T_{(\a\b|}{}^{g}T_{|\gamma)g}{}^{d}~
-~\nabla_{(\a|}T_{\b\g)}{}^{d}~=~0 \eea
 \bea
T_{(\a\b|}{}^{\la}T_{|\g)\la}{}^{\d}~-~T_{(\a\b|}{}^{g}T_{|\g)g}{}^{\d}
~-~\nabla_{(\a|}T_{|\b\g)}{}^{\d} -~\frac{1}{4}R^{(2)}{}_{(\a\b|
de}\s^{de}{}_{|\g)}{}^{\d} ~=~0 \eea \bea
T_{(\a\b|}{}^{\la}R{}_{|\g)\la}{}_{de} -T
_{(\a\b|}{}^{\la}R{}_{|\g)\la}{}_{de}
-~\nabla_{(\a|}R{}_{|\b\g)}{}_{de}~=~0\nn\\
\eea

To this list we now add the equations arising form the F sector

\bea \nabla_{[A}F_{|BC)}-T^{M}{}_{[AB}F_{M|C)}=0 \eea In our
previous work the super-current $A^{(1)}{}_{gef}$ was given by

\bea A^{(1)}{}_{gef}=+i\g \s_{gef}{}_{\epsilon \tau}T^{mn}{}^{
\epsilon}T_{mn}{}^{\tau} \eea

To begin with we modify $ A^{(1)}{}_{gef}$ to include matter
fields as in \cite{5}.

\bea A^{(1)}{}_{gef}\longrightarrow+i \s_{gef}{}_{\epsilon
\tau}[\g T^{mn}{} ^{\epsilon}T_{mn}{}^{\tau} +\b\lambda{}
^{\epsilon}\lambda{}^{\tau}] =A^{(\g)}{}_{gef}+A^{(\b)}{}_{gef}
\eea Here we use the notation where the superscript $\g$ or $\b$
is self evident. This has the effect of splitting the Bianchi
identities however we still have to be careful of cross terms. In
order to be cautious we will examine all contributions in the H
sector, in particular that for $H^{(2)}{}_{ a \b d}$, for
thoroughness (see appendix). In all of the Bianchi identities we
encounter the spinor derivative $\nabla_{\a}A^{(1)}{}_{abc}$. We
now show that a key equation which led to the previous second
order modified beta function favored ($\b FF$) solution is still
valid but with the modified $A^{(1)}{}_{abc}$.
 The spinor derivative of $A^{(1)}{}_{abc}$ will now contain the contributions due to the spinor derivative of $\lambda{}^{\tau}$ at zeroth order.
We have the following modifications from \cite{5}:

\bea
\nabla_{\g}T^{(0)}{}_{ef}{}^{\d}=-\frac{1}{4}\s^{mn}{}_{\g}{}^{\d}R^{(0)}{}_{efmn}
~+~T^{(0)}{}_{ef}{}^{\la}T^{(0)}{}_{\g\la}{}^{\d} \nn\\\eea

\bea
\nabla_{\g}\lambda^{(0)}{}^{\d}=-\frac{1}{4}\s^{mn}{}_{\g}{}^{\d}F_{mn}
~+~\la^{\la}T^{(0)}{}_{\g\la}{}^{\d} \nn\\\eea

The fundamental equation which enable a solution to be found with
higher order $\b FF$ constraints, now with the modified
supercurrent, $A^{(1)}{}_{pqr}$ is still given by

 \bea
T^{(0)}{}_{(\a\b|}{}^{\la}\s^{pqref}{}_{|\g)\la}A^{(1)}{}_{pqr}H^{(0)}{}_{def}-
\s^{pqref}{}_{(\a\b|}H^{(0)}{}_{def}\nabla_{|\g)}A^{(1)}{}_{pqr}\nn\\
=-24 \s^{g}{}_{(\a\b|}H^{(0)}{}_{d}{}^{ef}[\Omega^{(1)}{}_{|\g)
gef}]\eea

In the H sector we will have the following quantities \bea H_{ABC}=
H^{(\g\g)}{}_{ABC}+ H^{(\b\b)}{}_{ABC}+H^{(\b\g)}{}_{ABC}\eea

We will also adopt the modified torsion

\bea T^{(2)}{}_{\a\b}{}^{g}=-\frac{i\g}{6}\s^{pqref}{}_{\a\b}
H^{(0)}{}^{d}{}_{ef}A^{(1)}{}_{pqr}\eea

We then find no change in the form of the H sector results.

\bea \bar{H}^{(2)}{}_{\a\b d}=H^{(\g\b)}{}_{\a\b d}~
=~\s_{\a\b}{}^{g}[-~i\g
H^{(0)}{}_{def}A^{(\b)}{}_{g}{}^{ef}]\nn\\-
\s^{pqref}{}_{\a\b}[\frac{i\g}{12}H^{(0)}{}_{def}A^{(\b)}{}_{pqr}]
\eea

Equation (21) contains the second order contribution of the spinor
derivative  $\nabla_{\g} H_{\a\b g}$. Term by term we notice no
order $\b^2$ contributions occur. We also seek terms of the form
$O(\g\b)$ and note that none appear other than those resulting
from the modification to $A_{abc}$. We also find $H^{(2)}{}_{g \g
d}$ as in \cite{12}, but we write it in a different way

\bea
+\frac{i}{2}\s_{(\a\b|}{}^{g}H^{(2)}{}_{g|\g)d}=-\frac{\g}{12}\s^{\g|}{}_{(\a\b|}\s^{pqr}{}_{g}{}^{f}{}_{|\g)\la}A^{(1)}{}_{pqr}
T_{df}{}^{\la}\nn\\
+i\s^{g}{}_{(\a\b|}\{4\g
[\nabla_{|\g)}H^{(0)}{}_{d}{}^{ef}]L^{(1)}{}_{gef}+4\g
H^{(0)}{}_{d}{}^{ef}\nabla_{|\g)}L^{(1)}_{gef}\nn\\-\frac{\g}{2}[\nabla_{|\g)}H^{(0)}{}_{d}{}^{ef}]A^{(1)}{}_{gef}
-\frac{\g}{2}H^{(0)}{}_{d}{}^{ef}\nabla_{|\g)}A^{(1)}{}_{gef}-2\g
H^{(0)}{}_{g}{}^{ef}R^{(1)}{}_{|\g)def}
-2\g L^{(1)}{}_{g}{}^{ef}R^{(0)}{}_{|\g)def}\nn\\
+\frac{\g}{4}A^{(1)}{}_{gef}R^{(0)}{}_{|\g)def}
+2\g\nabla_{|\g)}[H^{(0)}{}_{d}{}
^{ef}H^{(0)}{}_{gef}]^{Order(1)}\}\nn\\~~~ \eea

The symmetrized result can be extracted as in \cite{5}. We list it
in appendix. We also believe that it can be simplified further. We
also have, along with (33) of the first reference \cite{12},

\bea i\s^{g}{}_{(\a\b|}T^{(2)}{}_{|\g) g}{}^{\d} =
 +2i\g\s^{g}{}_{(\a\b|} T^{(0)}{}^{ef}{}^{\d}
\Omega^{(1)}{}_{|\g)gef} \eea \bea
+~i\s^{g}{}_{(\a\b|}T^{(2)}{}_{|\g)gd}~=
+4i\g\s^{g}{}_{(\a\b|}H^{(0)}{}_{d}{}^{ef}[\Omega^{(1)}{}_{|\g)gef}]
+~\frac{\g}{6}\s^{g}{}_{(\a\b|}\s^{pqre}{}_{g}{}_{|\g)\f}A^{(1)}{}_{pqr}T^{(0)}{}_{de}{}^{\f}
\eea

\bea R_{\a\b de}
=-~\frac{i\g}{12}\s^{pqref}{}_{\a\b}A^{(1)}{}_{pqr}R_{ef de} \eea

\section{F Sector Bianchi Identities}

We have seen that in the H  and Torsion sectors the required minimal
Bianchi
 identities have been fully satisfied by simply modifying the super current as
 in equation (28). The fundamental identity $(21)$ is then used to solve in each case.
 This is the identity that enables such solutions to be obtained in the modified $\b$
  function favored set of constraints. Here we show that it can also be used to solve
  in the F sector coupled with an Ansatz.
To begin with we consider the following Bianchi Identities:

\bea
T_{(\a\b|}{}^{\la}F_{|\g)\la}~-~T_{(\a\b|}{}^{g}F_{|\g)}{}_{g}~
-~\nabla_{(\a|}F_{|\b\g)}~=~0 \eea \bea
\nabla_{\a}F_{ab}~-~\nabla_{[a|}F_{\a|b]}-T_{\a
[a|}{}^{M}F_{M}{}_{|b]} -T_{ab}{}^{M}F_{M}{}_{\a}~=~0 \eea We have
from \cite{5} to first order,  \bea
F^{(0)}{}_{\a\b}=F^{(1)}{}_{\a\b}=F^{(1)}{}_{\a d}=~0\eea \bea
F^{(0)}{}_{\a d}=-i\s_{d \a \f}\la^{\f}\eea

\bea
\nabla_{\a}F^{(0)}{}_{ef}=i\s_{[e|\a\f}\nabla_{|f]}\la^{\f}-2i\s^{g}{}_{\a\f}H^{(0)}{}_{gef}
\eea

Equation (29) is not satisfied by $F^{(2)}{}_{\a \b}=0$. We propose
the following candidate and show that it works. \bea
F^{(2)}{}_{\a\b}=-\frac{i\g}{12}\s^{pqref}{}_{\a\b}A^{(1)}{}_{pqr}F_{ef}\eea

At second order equation (29) becomes

\bea
T^{(0)}{}_{(\a\b|}{}^{\la}F^{(2)}{}_{|\g)\la}-\nabla_{(\a|}F^{(2)}{}_{|\b\g)}
-i\s^{g}{}_{(\a\b|}F^{(0)}{}_{|\g)}{}_{g}+
i\frac{\g}{6}\s^{pqref}{}_{(\a\b|}A^{(1)}{}_{pqr}H^{(0)}{}^{g}{}_{ef}[-i\s_{g |\g) \f}\la^{\f}]~~\nn\\
~=~0~~\eea or, more clearly \bea
T^{(0)}{}_{(\a\b|}{}^{\la}[-\frac{i\g}{12}\s^{pqref}{}_{|\g)\la}F_{ef}A^{(1)}{}_{pqr}]\nn\\
+\frac{i\g}{12}\s^{pqref}{}_{(\a\b|}F_{ef}\nabla_{|\g)}A^{(1)}{}_{pqr}
-~\frac{i\g}{12}\s^{pqref}{}_{(\a\b|}A^{(1)}{}_{pqr}\nabla_{|\g)}F_{ef}\nn\\
-~i\s^{g}{}_{(\a\b|}F^{(0)}{}_{|\g)}{}_{g}+
i\frac{\g}{6}\s^{pqref}{}_{(\a\b|}A^{(1)}{}_{pqr}H^{(0)}{}^{g}{}_{ef}[-i\s_{g |\g) \f}\la^{\f}]\nn\\
~=~0~~\eea Using (21) generates two solvable terms and we also set
up a cancellation. We also must use \bea
\s^{pqref}{}_{(\a\b|}\s_{e|\g)\f}=~-~\s^{pqref}{}_{\f(\a|}\s_{e|\b\g)}
\eea

Hence, we obtain

\bea
-i\s^{g}{}_{(\a\b|}F^{(2)}{}_{|\g)g}+2i\g\s^{g}{}_{(\a\b|}\Omega^{(1)}{}_{|\g)gef}F^{(0)}{}^{ef}
-\frac{\g}{6}\s^{g}{}_{(\a\b|}\s^{pqrgf}{}_{|\g)\f}A^{(1)}{}_{pqr}\nabla_{f}\la^{\f}\nn\\
+~\frac{i\g}{12}\s^{pqref}{}_{(\a\b|}A^{(1)}{}_{pqr}[-2i\s^{g}{}_{|\g)\f}H^{(0)}{}_{gef}]\nn\\
-i\frac{\g}{6}\s^{pqref}{}_{(\a\b|}A^{(1)}{}_{pqr}H^{(0)}{}^{g}{}_{ef}[-i\s_{g |\g) \f}\la^{\f}]
~=~0\nn\\~~\eea

The last two terms conveniently cancel. We find

\bea i\s^{g}{}_{(\a\b|}F^{(2)}{}_{|\g)
g}=+2i\g\s^{g}_{(\a\b|}\Omega^{(1)}{}_{|\g)gef}F^{(0)}{}^{ef}
-\frac{\g}{6}\s^{g}{}_{(\a\b|}\s^{pqrgf}{}_{|\g)\f}A^{(1)}{}_{pqr}\nabla_{f}\la^{\f}\eea
or \bea F^{(2)}{}_{\g g}=+2\g \Omega^{(1)}{}_{\g gef}F^{(0)ef}
+i\frac{\g}{6}\s^{pqr}{}_{g}{}^{f}{}_{\g
\f}A^{(1)}{}_{pqr}\nabla_{f}\la^{\f}\eea

 Finally for completeness, we consider the derivatives, $\nabla_{\a}F^{(2)}{}_{bc}$
 and $\nabla_{(\a|}F^{(2)}{}_{|\b)g}$.  We have

  \bea
  \nabla_{(\a|}F_{|\b)d}-T_{\a\b}{}^{g}F_{gd}-T_{\a\b}{}^{\la}F_{_|la d}+T_{(\a|d}{}^{g}F_{g|\b)}+T_{(\a|d}{}^{\la}F_{\la|\b)}=0\nn\\
  \eea
 and
   \bea
  \nabla_{\a}F_{bc} = \nabla_{[ b|}F_{\a|c]}+T_{\a[b}{}^{g}F_{g|c]}+T_{\a[b|}{}^{\la}F_{\la |]}-T_{bc}{}^{g}F_{g \a}+T_{bc}{}^{\la}F_{\la \a}\nn\\
  \eea

 Substitution of (31), (32), (33), (34) and (40), as well as other results quoted in \cite{5}
 and
 \bea T^{(2)}{}_{\a\b}{}^{\la}=-\frac{i\g}{12}\s^{pqref}{}_{\a\b}
A^{(1)}{}_{pqr}T^{(0)}{}^{\la}{}_{ef}\eea give, respectively for
(41) and (42), \bea
  -i\s_{d(\a|\f}\nabla_{|\b)}\la^{\f}|^{(2)}+2\g\nabla_{(\a|}\Omega^{(1)}{}_{|\b)def}F^{ef}
  +4i\g\Omega^{(1)}{}_{(\a| d}{}^{ef}\s_{e|\b)\f}\nabla_{f}\la^{\f}\nn\\
   -4i\g\Omega^{(1)}{}_{(\a| d}{}^{ef}\s_{g|\b)\f}H^{(0)}{}^{g}{}_{ef}
  +\frac{i\g}{6}\s^{pqr}{}_{d}{}^{f}{}_{(\a|\f}(\nabla_{|\b)}A^{(1)}{}_{pqr})[\nabla_{f}\la^{\f}]\nn\\
+\frac{i\g}{6}\s^{pqr}{}_{d}{}^{f}{}_{(\a|\f}A^{(1)}{}_{pqr}[\nabla_{|\b)}\nabla_{f}\la^{\f}]
-i\s^{g}{}_{\a\b}F^{(2)}{}_{gd}+\frac{\g}{12}\s^{pqref}{}_{\a\b}\s_{d\la\f}A^{(1)}{}_{pqr}T_{ef}{}^{\la}\la^{\f}\nn\\
-2\g T^{(0)}{}_{\a\b}{}^{\la}\Omega^{(1)}{}_{\la
def}F^{(0)}{}^{ef}
-i\frac{\g}{6}T^{(0)}{}_{\a\b}{}^{\la}\s^{pqr}{}_{d}{}^{f}{}_{\la
f}A^{(1)}{}_{pqr}\nabla_{f}\la^{\f}
-i\s_{g(\a|\f}\la^{\f}T^{(2)}{}_{|\b)d}{}^{g} =0 \eea

\bea \nabla_{\a}F_{bc}=2\g\nabla_{[b|}\Omega^{(1)}{}_{\a |c]
ef}F^{ef}
  -i\frac{\g}{6}\s^{pqr}{}_{[b|}{}^{f}{}_{\a \f}A^{(1)}{}_{pqr}\nabla_{|c]}\nabla_{f}\la^{\f}\nn\\
  -i\frac{\g}{6}\s^{pqr}{}_{[b|}{}^{f}{}_{\a \f}[\nabla_{|c]}A^{(1)}{}_{pqr}]\nabla_{f}\la^{\f}\nn\\
  +T^{(2)}{}_{\a[b|}{}^{g}F^{(0)}{}_{g|c]}
  -iT^{(2)}{}_{\a[b|}{}^{\la}\s_{|b]\la\f}\la^{\f}+2\g T^{(0)}{}^{g}{}_{bc}\Omega_{\a gef}F^{ef}\nn\\
  +i\frac{\g}{6} \s^{pqr}{}_{g}{}^{f}{}_{\la \f} A^{(1)}{}_{pqr}T^{(2)}{}_{bc}{}^{g} \nabla_{f}\la^{f}
  -i\g_{g\a\f}\la^{\f}T^{(2)}{}_{bc}{}^{g}
  -i\frac{\g}{12}\s^{pqref}{}_{\la\a}A^{(1)}{}_{pqr}F^{(0)}{}_{ef}T^{(0)}{}_{bc}{}^{\la}
\eea

\section{Modified Commutator}
Finally, we note how the commutator (4) is modified. We have

\bea
[\nabla_{\a},\nabla_{\b}\}=-~\frac{i\g}{12}\s^{pqref}{}_{\a\b}A^{(1)}{}_{pqr}[T_{ef}{}^{\g}\nabla_{\g}
+2
H^{(0)}{}_{ef}{}^{g}\nabla_{g}+\frac{1}{2}R^{(0)}{}_{ef}{}^{mn}M_{mn}+iF^{(0)}{}_{ef}{}^{I}t_{I}]\nn\\~
\eea

But \bea 2 H^{(0)}{}_{ef}{}^{g} = -T^{(0)}{}_{ef}{}^{g},~~~~~ \eea
And \bea R^{(0)}{}_{ef}{}^{mn}M_{mn} =
-\frac{1}{8}R^{(0)}{}_{ef}{}_{\la}{}^{\d}\s^{mn}{}_{\d}{}^{\la}\eea
so  we get the modified geometry proportional to $\s^{5}$ as
follows,

\bea
[\nabla_{\a},\nabla_{\b}\}|^{(2)}=-~\frac{i\g}{12}\s^{pqref}{}_{\a\b}A^{(1)}{}_{pqr}[T{}^{(0)}{}_{ef}{}^{\g}\nabla_{\g}
- T^{(0)}{}_{ef}{}^{g}\nabla_{g}\nn\\-
\frac{1}{16}R^{(0)}{}_{ef}{}_{\la}{}^{\d}\s^{mn}{}_{\d}{}^{\la}M_{mn}
-iF^{(0)}{}_{ef}{}^{I}t_{I}] \eea

A point conceptually important can be stressed here by observing
that the latter expression agrees in its structure with the
conjecture in Ref. [5] for supersymmetric Yang-Mills couplings.

 \section{Conclusion}

 The geometrical methods currently known as deformations \cite{1} and the constraints
often referred to as beta function favored constraints
\cite{vash,conf2} allowed for the determination of the most
general higher derivative Yang-Mills action to the third order,
which is globally supersymmetric and Lorentz covariant in D=10
spacetime (see e.g. \cite{more}), a result which is important for
topologically nontrivial gauge configurations of the vector field,
such as, for instance, in the case of compactified string theories
on manifolds with topologically nontrivial properties. Building
upon our previous works on the subject, we have been able here to
provide a consistent inclusion of Yang Mills matter fields into
string corrected (deformed), D=10, N=1 Supergravity. Our solution
to the Bianchi identities, obtrained within the framework of the
modified beta function favored constraints, holds to the second
order in the string slope parameter $\g$ and includes also the
Yang Mills fields. We obtained as well a consistent
  solution in the torsion, curvature and H sectors with a Yang Mills modified supercurrent $A_{abc}$.
  Following a technique we developed in earlier papers, we also found a solution in the F sector and
  gave an explicit formula for the modification induced in the commutator expression.

\subsection*{Acknowledgements}
This work of has been supported in part by the European Community
Human Potential Program under contract MRTN-CT-2004-005104
\textit{``Constituents, fundamental forces and symmetries of the
universe''}.

\section{Appendix}
For the sake of extra caution as the Bianchi identity for example
for $H_{\a b d}$ is long, we can write out the full version of as
follows.

\bea \frac{1}{2}\nabla_{(\a|}H_{|\b\g)d}{}^{(\g\g; \b\b; \g
\b)}-\nabla_{d}H_{\a\b\g}{}^{(Order 2)}
-\frac{1}{2}T^{(0)}{}_{(\a\b|}{}^{\la}H^{(\g\g)}{}_{\la|\g)d}-
\frac{1}{2}T^{(0)}{}_{(\a\b|}{}^{\la}H^{(\g\b)}{}_{\la|\g)d}\nn\\
-\frac{1}{2}T^{(0)}{}_{(\a\b|}{}^{\la}H^{(\b\b)}{}_{\la|\g)d}
-\frac{1}{2}T^{(\g)}{}_{(\a\b|}{}^{\la}H^{(\b)}{}_{\la|\g)d}
-\frac{1}{2}T^{(\g)}{}_{(\a\b|}{}^{\la}H^{(\g)}{}_{\la|\g)d}
-\frac{1}{2}T^{(\b)}{}_{(\a\b|}{}^{\la}H^{(\b)}{}_{\la|\g)d}\nn\\
-\frac{1}{2}T^{(\b)}{}_{(\a\b|}{}^{\la}H^{(\g)}{}_{\la|\g)d}
-\frac{1}{2}T^{(\g\g)}{}_{(\a\b|}{}^{\la}H^{(0)}{}_{\la|\g)d}
-\frac{1}{2}T^{(\g\b)}{}_{(\a\b|}{}^{\la}H^{(0)}{}_{\la|\g)d}
-\frac{1}{2}T^{(\b\b)}{}_{(\a\b|}{}^{\la}H^{(0)}{}_{\la|\g)d}\nn\\
-\frac{1}{2}T^{(0)}{}_{(\a\b|}{}^{g}H^{(\g\g)}{}_{g|\g)d}-\frac{1}{2}T^{(0)}{}_{(\a\b|}{}^{g}H^{(\g\b)}{}_{g|\g)d}-
\frac{1}{2}T^{(0)}{}_{(\a\b|}{}^{g}H^{(\b\b)}{}_{g|\g)d}
-\frac{1}{2}T^{(\g)}{}_{(\a\b|}{}^{g}H^{(\b)}{}_{g|\g)d}\nn\\
-\frac{1}{2}T^{(\b)}{}_{(\a\b|}{}^{g}H^{(\g)}{}_{g|\g)d}
+\frac{1}{2}T^{(\g\g)}{}_{d(\a|}{}^{g}H^{(0)}{}_{g|\b\g)}+\frac{1}{2}T^{(\g\b)}{}_{d(\a}{}^{g}H^{(0)}{}_{g|\b\g)}\nn\\
+\frac{1}{2}T^{(\b\b)}{}_{d(\a|}{}^{g}H^{(0)}{}_{g|\b\g)}
+\frac{1}{2}T^{(\g)}{}_{d(\a|}{}^{g}H^{(\g)}{}_{g|\b\g)}+\frac{1}{2}T^{(\g)}{}_{d(\a}{}^{g}H^{(\b)}{}_{g|\b\g)}
+\frac{1}{2}T^{(\b)}{}_{d(\a|}{}^{g}H^{(\g)}{}_{g|\b\g)}+\nn\\
+\frac{1}{2}T^{(\b)}{}_{d(\a|}{}^{g}H^{(\b)}{}_{g|\b\g)}+\frac{1}{2}T^{(0)}{}_{d(\a|}{}^{g}H^{(\g\g)}{}_{g|\b\g)}
+\frac{1}{2}T^{(0)}{}_{d(\a|}{}^{g}H^{(\g\b)}{}_{g|\b\g)}+\frac{1}{2}T^{(0)}{}_{d(\a|}{}^{g}H^{(\b\b)}{}_{g|\b\g)}\nn\\
+{\g}[R^{(\g)}{}_{(\a\b|ef}R^{(0)}{}_{|\g)d}{}^{ef}+R^{(\b)}{}_{(\a\b|ef}R^{(0)}{}_{|\g)d}{}^{ef}+
R^{(0)}{}_{(\a\b|ef}R^{(\b)}{}_{|\g)d}{}^{ef}+R^{(0)}{}_{(\a\b|ef}R^{(\g)}{}_{|\g)d}{}^{ef}]\nn\\+
{\b}[F^{(\g)}{}_{(\a\b|}F^{(0)}{}_{|\g)d}+F^{(\b)}{}_{(\a\b|}F^{(0)}{}_{|\g)d}+
F^{(0)}{}_{(\a\b|}F^{(\b)}{}_{|\g)d}+F^{(0)}{}_{(\a\b|}F^{(\g)}{}_{|\g)d}]=0~\nn\\~~~\eea

Applying the first order constraints and allowing $H_{ABC}$ terms to
drop out leaves the  $\bar{H}_{ABC}$ contributions.

\bea \frac{1}{2}\nabla_{(\a|}H_{|\b\g)d}{}^{( \b\b; \g \b)}-
\frac{1}{2}T^{(0)}{}_{(\a\b|}{}^{\la}H^{(\g\b)}{}_{\la|\g)d}
-\frac{1}{2}T^{(0)}{}_{(\a\b|}{}^{\la}H^{(\b\b)}{}_{\la|\g)d}\nn\\
-\frac{1}{2}T^{(\g\b)}{}_{(\a\b|}{}^{\la}H^{(0)}{}_{\la|\g)d}
-\frac{1}{2}T^{(\g\b)}{}_{(\a\b|}{}^{\la}H^{(0)}{}_{\la|\g)d}
-\frac{1}{2}T^{(0)}{}_{(\a\b|}{}^{g}H^{(\g\b)}{}_{g|\g)d}\nn\\
-\frac{1}{2}T^{(0)}{}_{(\a\b|}{}^{g}H^{(\b \b)}{}_{g|\g)d}
+\frac{1}{2}T^{(\g\b)}{}_{d(\a}{}^{g}H^{(0)}{}_{g|\b\g)}
+\frac{1}{2}T^{(\b\b)}{}_{d(\a|}{}^{g}H^{(0)}{}_{g|\b\g)}\nn\\
+\frac{1}{2}T^{(\g\g)}{}_{d(\a|}{}^{g}H^{(0)}{}_{g|\b\g)}
+{\g}[R^{(\b)}{}_{(\a\b|ef}R^{(0)}{}_{|\g)d}{}^{ef}+
R^{(0)}{}_{(\a\b|ef}R^{(\b)}{}_{|\g)d}{}^{ef}] =0\eea

We therefore find as before that no order $\b^{2}$ terms exist. Also
there is no need to make modifications other than the adjustment of
$A_{abc}$ in the H sector.

\end{document}